\documentclass{article}

\usepackage{arxiv}
\usepackage{amsmath}
\usepackage[utf8]{inputenc} % allow utf-8 input
\usepackage[T1]{fontenc}    % use 8-bit T1 fonts
\usepackage{hyperref}       % hyperlinks
\usepackage{url}            % simple URL typesetting
\usepackage{booktabs}       % professional-quality tables
\usepackage{amsfonts}       % blackboard math symbols
\usepackage{nicefrac}       % compact symbols for 1/2, etc.
\usepackage{microtype}      % microtypography
\usepackage{lipsum}
\usepackage{graphicx}
\graphicspath{ {./images/} }

\title{Evolutionary origin of the bipartite architecture of dissipative cellular networks}

\author{
 Bowen Shi \\
  Institute for Advanced Study in Physics\\
  Zhejiang University\\
  Hangzhou 310058, China\\
  \texttt{sbw1996@zju.edu.cn} \\
   \And
 Long Qian \\
  Center for Quantitative Biology\\
  Academy for Advanced Interdisciplinary Studies\\
  Peking University\\
  Beijing 100871, China\\
  \texttt{long.qian@pku.edu.cn} \\
  \And
 Qi Ouyang \\
  Institute for Advanced Study in Physics\\
  Zhejiang University\\
  Hangzhou 310058, China\\
  \texttt{qi@zju.edu.cn} \\
}

\begin{document}
\maketitle
\begin{abstract}
Energy dissipation underlies biological functions that operate far from equilibrium, yet the topological principles by which reaction networks organize energetic input remains poorly understood. Here, we combine evolutionary simulations with theoretical theory to investigate how dissipative architecture evolves under selection for function. Using canonical motifs for kinetic proofreading, biochemical oscillation, and adaptive response, we allow reaction topologies to mutate while selection acts only on functional performance. Across motifs, networks reproducibly evolve to a bipartite architecture in which high-energy fuel molecules are topologically decoupled from functional substrate-conversion motifs and assigned to dedicated driving cycles. This outcome emerges without direct selection for dissipation, yet it enables greater energy utilization and improves robustness to kinetic and structural perturbations. Analytical solutions of a minimal proofreading model demonstrate that direct fuel-substrate coupling imposes an upper bound on non-equilibrium driving, whereas fuel decoupling is a necessary condition for the system to approach maximal non-equilibrium driving and functional accuracy. Extending these findings to a composite network composed of coupled sub-functional modules, we show that fuel decoupling robustly emerges and is further promoted by modular evolutionary history. These results establish fuel decoupling and dedication as an evolutionary organizing principle for dissipative cellular networks, providing a physical rationale for the separation between cellular energy currencies and the functional modules they power.
\end{abstract}

% keywords can be removed

\section{Introduction}
\begin{figure}%[tbhp]
	\centering
	\includegraphics[width=.8\linewidth]{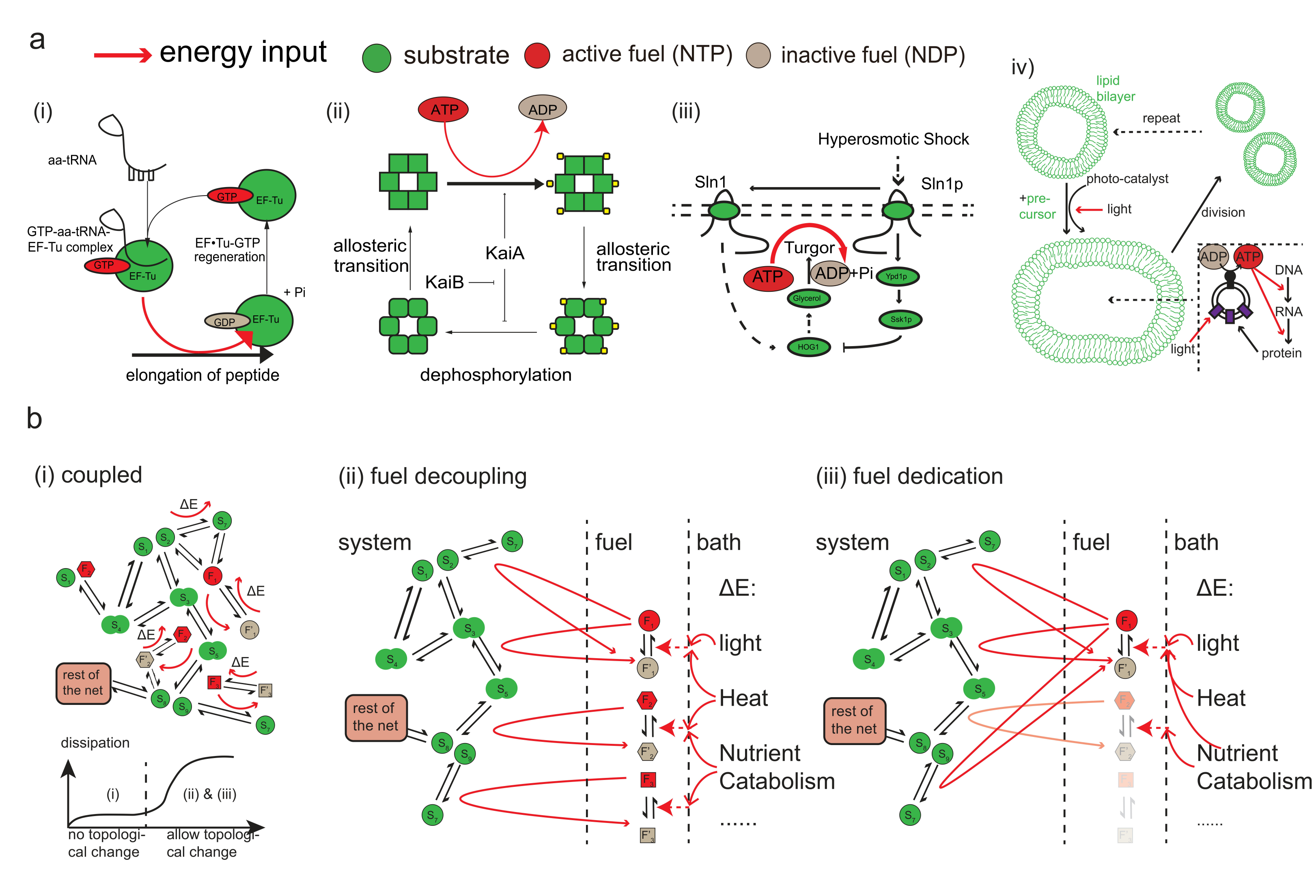}
	\caption{
	\textbf{Conceptual demonstration of fuel decoupling and dedication.}
	\textbf{(a)} Representative non-equilibrium biological and prebiotic motifs driven by external energy input (red arrows). 
	\textbf{(i)} Elongation of a protein polymer in mRNA translation and the corresponding kinetic proofreading model~\cite{hopfield1974}; 
	\textbf{(ii)} Circadian oscillation by KaiC and its simplified activator--inhibitor oscillatory model~\cite{cao2015}, yellow dots represent the phosphorylated state of KaiC; 
	\textbf{(iii)} Osmotic adaptation by Sln1/Sln2 and the abstract adaptive response model~\cite{lan2012}; 
	\textbf{(iv)} A synthetic minimal cell model using a photo-catalyst and visible light as the energy source to replicate liposome vesicles~\cite{declue2009}. 
	\textbf{(b)} Schematic illustration of network evolution under different topological constraints. Substrates and fuel molecules, as well as energy inputs, are represented in the same symbols as in (a). 
	\textbf{(i)} When the reaction topology is fixed, the system remains in a stable dissipative state with limited energy flux. The lower panel shows the potential increase in maximal achievable dissipation when a static topology (state i) is allowed to undergo topological evolution (states ii and iii). 
	\textbf{(ii)} Schematic for fuel decoupling, where energetically rich fuels become isolated from direct substrate interconversions and are solely used for transducing energy from the external heat bath. 
	\textbf{(iii)} Schematic for fuel dedication, in which energy input from diverse sources (light, heat, nutrient catabolism) converges onto a few specialized fuel molecules (e.g., ATP and GTP).
	}
	\label{fig:fig1}
\end{figure}

Biological systems are inherently dissipative, with continuous energy input playing a crucial role in a wide range of essential cellular functions such as error correction, temporal coordination, and adaptive responses~\cite{eisenberg1985,hill1989,beard2008} (Fig.~\ref{fig:fig1}a). For instance, kinetic proofreading consumes energy to enhance molecular specificity beyond the equilibrium limit based on cognate and non-cognate substrate binding differences~\cite{hopfield1974}. Biochemical oscillations depend on sustained energy supply to maintain coherent dynamics and improve temporal precision~\cite{cao2015}. Cellular signal transduction pathways also rely on energy consumption to generate finely controlled adaptive behaviors for robust decision-making~\cite{lan2012,ma2009}. Moreover, free energy is indispensable for facilitating the rapid turnover rates and irreversibility characteristic of biomolecular self-assembly~\cite{marsland2018}. Together, these systems illustrate a general feature of living matter: functional performance in biological networks is tightly coupled to non-equilibrium driving.

In parallel, synthetic living systems, sometimes referred to as protocells, have been constructed around metabolic reactions~\cite{rasmussen2016,adamski2020,xu2016,declue2009,yu2001,berhanu2019}. For instance, DeClue et al. utilized a light-driven ruthenium tris(bipyridine) complex to catalyze redox reactions on precursor amphiphilic molecules, enabling self-replicative micro-vesicle growth. Similarly, Berhanu et al. reconstituted a synthetic central dogma fueled by light-driven ATP synthase~\cite{berhanu2019}.

Understanding how such dissipative biological networks originate and organize remains a key challenge. Previous studies have approached this problem from complementary but incomplete perspectives. Wołos et al. interrogated the formation of primitive biological networks by systematically exploring non-enzymatic pathways in a prebiotic environment. Their findings suggested the spontaneous emergence of critical biomolecules and characteristic pathway modules in biological networks~\cite{wolos2020}. However, due to computational limitations, their simulations excluded external energy inputs. On the other hand, England et al. proposed a generic model of dissipative networks wherein a random chemical reaction network acquires external energy by coupling select reactions to specific ``fuel'' species, resulting in extreme net energy fluxes on these highly irreversible reactions (Fig.~\ref{fig:fig1}b(i)). They demonstrated that such systems spontaneously align toward rare but highly dissipative states, indicating the emergence of dissipative networks through dynamic state evolution~\cite{horowitz2017}. However, the topology of the network, i.e., the whole set of reaction linkages, remained unchanged during this dynamic process. In contrast, biological networks undergo constant evolutionary modifications to their reaction topology through mutation and selection. How network topology itself evolves under energetic driving, and whether certain topological organizations are favored by non-equilibrium functional demands, remains largely unexplored.

Strikingly, real biological networks display a highly conserved architectural organization with respect to energy usage, which manifests in two distinctive features. First, fuel dedication (Fig.~\ref{fig:fig1}b(iii)). Despite the diversity of available energy sources, including light, redox gradients, and nutrient catabolism~\cite{rasmussen2016,adamski2020,xu2016,declue2009,yu2001,berhanu2019}, cellular energy input converges onto a small number of dedicated fuel molecules, most prominently adenosine triphosphate (ATP). ATP synthesis by ATP synthase is facilitated by oxidative phosphorylation on the mitochondrial inner membrane, as well as in the photosynthetic electron transport chain on the thylakoid membranes.

Second, fuel decoupling (Fig.~\ref{fig:fig1}b(ii)). Apart from a few mechanisms pertaining to the global regulation of cellular physiological states---such as the PKA pathways and stringent response pathways~\cite{sassone2012,irving2021}, which necessitate the conversions from ATP to cAMP and from GTP to (p)ppGpp---ATP and similar fuels rarely directly engage in substrate interconversion within signaling, regulatory and decision-making networks. Instead, they primarily serve as the ``energy currency'' that biases reaction cycles away from equilibrium. This separation between energetic input and functional matter flux gives rise to a bipartite network architecture that is ubiquitous in modern cells.

The prevalence of this bipartite organization raises a fundamental question. Is decoupling fuels from functional motifs necessary for achieving strong non-equilibrium driving and high functional performance, or does it simply represent one of many possible energetic implementations? Here, we hypothesize that dissipative topology marks an evolved feature of biological networks, ensuring functional accuracy by maintaining a stable energy input through dedicated fuel molecules that are topologically decoupled from regulatory processes.

To investigate this hypothesis and explore the origin of robust and highly dissipative network topology in cells, we conducted evolutionary simulations of key functional motifs subject to energy dissipation. We found that, even without explicit selection on energy dissipation, networks evolved toward configurations in which the highest-energy molecule became decoupled from functional motifs. Furthermore, this decoupling enhanced the system's robustness against parameter and structural perturbation. We further complement these simulations with analytical treatment of minimal models to identify the physical constraints linking topology, dissipation, and non-equilibrium driving. Finally, by extending our analysis to a toy system combining self-replication and oscillatory control to mimic primitive cell replication, we highlight the generality of this organizational principle underlying biological networks' energetic architecture.

\section{Results}

\subsection{Fuel decoupling and dedication as an evolved feature under functional selection}

\begin{figure}%[tbhp]
	\centering
	\includegraphics[width=.8\linewidth]{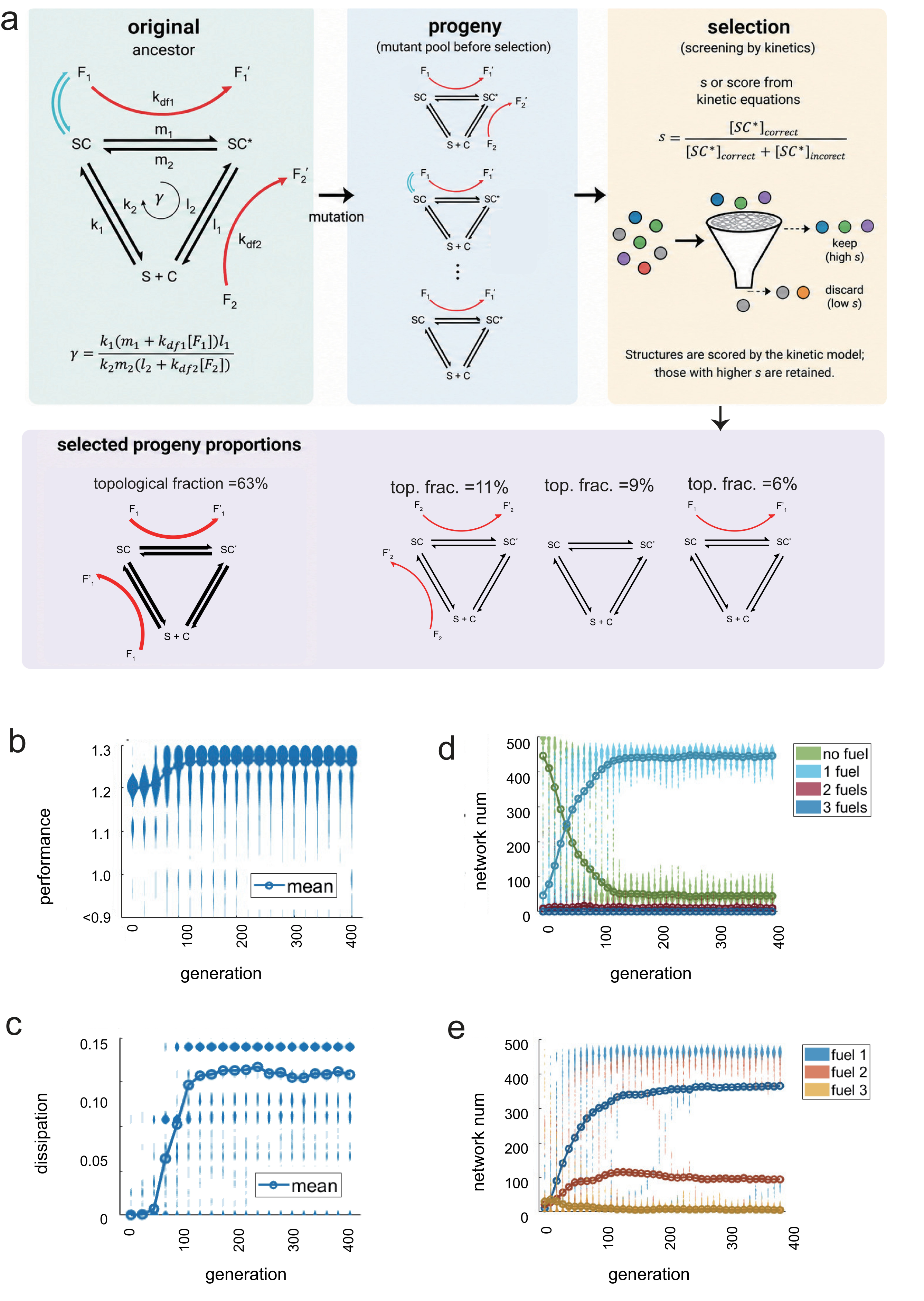}
	\caption{
	\textbf{Evolutionary simulation of the proofreading model.}
	\textbf{(a)} Schematic of the evolutionary simulation. Each network topology originates from a randomly generated ancestor and undergoes structural mutations (addition, deletion, or reversal of reactions). After scoring by functional performance, high-fitness networks are selected to seed the next generation. The score, defined as the ratio of correctly incorporated substrates in the case of kinetic proofreading, determines the selective advantages. 
	\textbf{(b)} Performance score distribution of an evolving population (size = 500) in a representative simulation, showing the progressive increase of network performance during evolution. 
	\textbf{(c)} Network energy dissipation distribution of the same evolving population as in \textbf{(b)}, which increases despite dissipation not being directly selected for. 
	\textbf{(d)} Evolution of fuel--substrate separation across 23 independent simulations. Colors denote the number of fuels that became decoupled (0--3). Open circled lines represent the mean number of networks in each group. 
	\textbf{(e)} Evolution of decoupled fuel types. Each color represents a distinct fuel ($F_1$--$F_3$), open circled lines represent the mean number of networks in each group; the results show a preference for decoupling the highest-energy fuel $F_1$.
	}
	\label{fig:fig2}
\end{figure}

To investigate whether dissipative topologies can evolve from generic biochemical networks composed of interconnected fuels and regulatory molecules, we performed evolutionary simulations of a generalized Wright--Fisher framework on three classic non-equilibrium biochemical networks (Fig.~\ref{fig:fig2}a and Fig.~S1): kinetic proofreading of RNA biosynthesis, adaptive responses in sensory signal transduction, and biomolecule oscillations~\cite{hopfield1974,cao2015,lan2012}. In all cases, networks consist of regulatory substrates (e.g., $S$ and $C$ in Fig.~\ref{fig:fig2}a) interconverting under mass and energy conservation, and three fuel molecules ($F_1$, $F_2$, and $F_3$) supplied at a constant level from external sources. Each fuel molecule exists in high- and low-energy states ($F_i$ and $F_i'$, respectively), analogous to nucleoside triphosphate/diphosphate pairs. We assigned the energy gaps $\Delta E_{F_i} = E_{F_i} - E_{F_i'}$ to the three fuel molecules in decreasing order $\Delta E_{F_1} > \Delta E_{F_2} > \Delta E_{F_3}$.

In all stages of the simulations, fuel molecules can either participate directly in a substrate interconversion reaction $F \rightleftharpoons S$ with reaction rates $k_+/k_- \sim \exp(E_F - E_S)$, or drive dissipative reactions $SC + F \rightleftharpoons SC^* + F'$ with reaction rates $k_+/k_- = \exp(E_{SC} - E_{SC^*} + \Delta E_F)$. In contrast, regulatory molecules are only able to participate in substrate interconversion (e.g., $S + C \rightleftharpoons SC$). Given the vast topological search space (Table~1), each evolutionary run was initialized with random topologies containing the core motif among other edges to ensure basic functionality (Methods and Supplementary Note~2). During each generation, three types of topological mutations were allowed at equal probabilities (1\% per edge): (i) addition or deletion of a substrate-conversion reaction, (ii) addition or deletion of a dissipative reaction, and (iii) reversal of dissipative reaction directionality (Fig.~\ref{fig:fig2}a and Fig.~S4). The fitness of each topology was determined by its numerically simulated functional performance. For example, kinetic proofreading topologies were evaluated by the selective pressure $s$, defined as the ratio of correctly incorporated monomers (Fig.~\ref{fig:fig2}a, Supplementary Note~2, and Methods). Dissipation was quantified as $W = \sum_{\text{all reactions}} (\eta_{\mathrm{forward}} - \eta_{\mathrm{backward}})\ln\left(\eta_{\mathrm{forward}}/\eta_{\mathrm{backward}}\right)$, where $\eta$ denotes the matter flux within each reaction cycle.
\begin{table}[t]
\centering
\caption{\textbf{Table 1. Topological parameters of simulated networks}}
\label{tab:topology}
\begin{tabular}{|l|c|c|c|}
\hline
\textbf{Core Motif} &
\textbf{Topological space size} &
\textbf{Mutation rate (per site)} &
\textbf{Number of mutation sites} \\
\hline

Proofreading &
$373248 = (3^3 \cdot 3^3)2^6 2^3$ &
0.01 &
$15 = (3+3)+6+3$ \\
\hline

Activator--Inhibitor (A--I) &
$214990848 = (3^4 \cdot 3^4)2^{15}2^0$ &
0.01 &
$23 = (4+4)+15+0$ \\
\hline

Incoherent Feedforward Loop (IFFL) &
$2985984 = (3^3 \cdot 3^3)2^{12}2^0$ &
0.01 &
$15 = (3+3)+12+0$ \\
\hline

Negative Feedback Loop (NFBL) &
$2985984 = (3^3 \cdot 3^3)2^{12}2^0$ &
0.01 &
$15 = (3+3)+12+0$ \\
\hline

\end{tabular}
\end{table}
We first focused on the kinetic proofreading network, which provides a minimal and well-characterized model of energy-dependent error correction~\cite{hopfield1974}. Starting from randomly generated topologies containing the core proofreading motif, evolutionary selection led to a progressive increase in assembly accuracy over generations (Fig.~\ref{fig:fig2}b). Although energy dissipation was not directly included in the fitness function, total network dissipation increased concomitantly along evolutionary trajectories (Fig.~\ref{fig:fig2}c). Dissipation $W$ closely tracked improvements in functional performance $s$. This correlation is consistent with previous analyses linking proofreading accuracy to non-equilibrium driving~\cite{hopfield1974,qian2006}, but here arises as an emergent consequence of topological evolution.

Strikingly, the increase in energy dissipation was accompanied by systematic changes in network architecture. As evolution proceeded, a growing proportion of networks exhibited topological separation between fuel molecules and the core functional motif, such that at least one fuel no longer participated directly in substrate interconversion reactions (Fig.~\ref{fig:fig2}d). While topologies in which multiple fuel molecules were decoupled occasionally appeared, these configurations were transient and remained rare. Instead, the dominant evolutionary outcome consisted of networks in which exactly one fuel molecule became decoupled from the functional core.

Analysis of fuel identity revealed a strong bias in this decoupling process. The fuel molecule with the largest energy gap, corresponding to the highest-energy fuel $F_1$, was preferentially decoupled, whereas lower-energy fuels ($F_2$ and $F_3$) were decoupled only sporadically and failed to persist over evolutionary time (Fig.~\ref{fig:fig3}b,c). This pattern suggests that functional selection alone consistently drives networks toward architectures in which the strongest energetic input is topologically decoupled from the regulatory reaction motif. Meanwhile, utilizing energetically inferior fuels provides little selective advantage for enhancing the performance of evolving networks.

Importantly, this evolutionary pattern is not specific to proofreading. Parallel simulations of oscillatory networks and two adaptive response motifs yielded qualitatively similar outcomes: functional optimization was consistently accompanied by increased dissipation (Fig.~S10--S13 and Methods). Despite their drastically different topologies and functions, all systems exhibited the spontaneous emergence of fuel decoupling and dedication as robust evolutionary outcomes of functional selection, rather than a consequence of explicit energetic optimization or model-specific assumptions.

\subsection{Preferential decoupling of the strongest fuel maximizes dissipation and performance}

\begin{figure}%[tbhp]
	\centering
	\includegraphics[width=.8\linewidth]{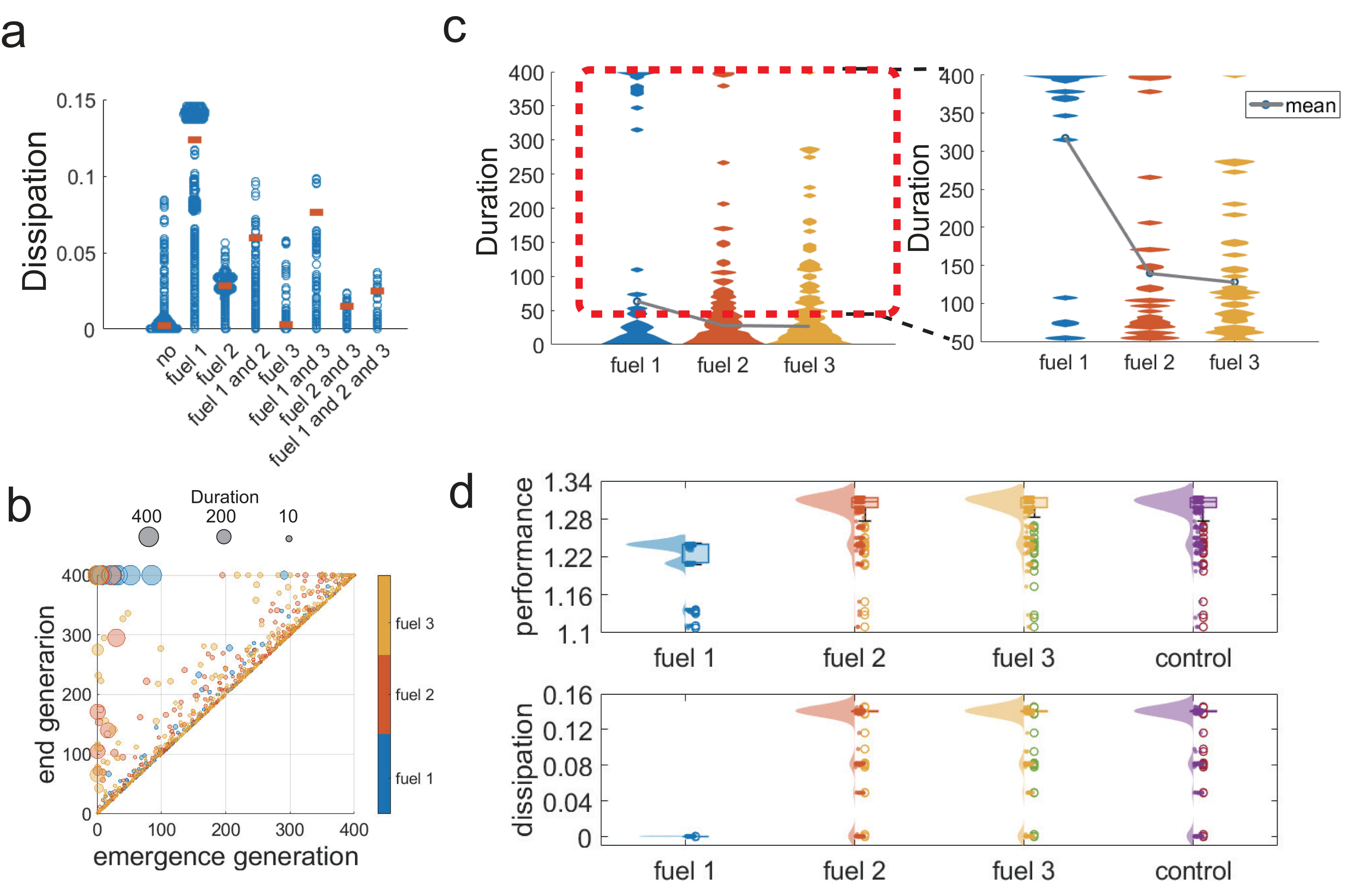}
	\caption{
	\textbf{Preferential fuel-decoupling in evolutionary simulations.}
	\textbf{(a)} Relationship between fuel decoupling state and total network dissipation. The x-axis indicates the identity of fuels that became decoupled, and the y-axis shows the corresponding dissipation values of individual networks selected from 23 independent evolutionary simulations. Each circle represents one evolved network. Orange bars represent mean dissipation of each group. 
	\textbf{(b)} The duration of fuel-decoupling events across 23 independent simulations. Each circle represents a decoupling event with the x-axis indicating the generation at which it first appears and the y-axis indicating the generation at which it disappears; color denotes fuel identity as shown in the color bar. Circle size indicates the duration each decoupling event persisted in the simulations. 
	\textbf{(c)} Duration of decoupling events for each fuel type. The y-axis indicates the number of generations over which each decoupled state persisted, and the width of each distribution corresponds to the number of networks exhibiting that state. Right panel highlights events persisting for more than 50 generations (red dashed box in the left panel). Open circled lines represent the mean durations in each group. 
	\textbf{(d)} Effect of fuel recoupling on evolved networks. The y-axis shows performance (upper panel) and dissipation (lower panel) of recoupled networks. Distributions are represented by violin plots (left) and box plots (right) for each recoupling condition.
	}
	\label{fig:fig3}
\end{figure}

To understand why evolutionary dynamics consistently favor the decoupling of a single fuel molecule, we analyzed the dissipation and performance of evolved network topologies classified by their fuel--substrate decoupling states. Across all network topologies obtained from simulations, total energy dissipation exhibited a strong dependence on which fuels were decoupled and to what extent (Fig.~\ref{fig:fig3}a). Networks in which the highest-energy fuel---and only the highest-energy fuel---($F_1$) was decoupled displayed substantially higher dissipation than networks with decoupling of weaker fuels. In contrast, networks in which multiple fuels were simultaneously decoupled, or in which fuel and substrate species remained extensively mixed, showed markedly reduced dissipation, indicating inefficient utilization of external energy input.

To assess the evolutionary stability of these decoupling states, we examined the temporal persistence of fuel-decoupled topologies across independent evolutionary trajectories. For each simulation, we recorded the generation at which a given fuel-decoupling event first appeared and the generation at which it was subsequently lost, if at all (Fig.~\ref{fig:fig3}b and Fig.~S10--S13). Decoupling of the strongest fuel molecule ($F_1$), once established, almost invariably persisted until the end of evolution, regardless of whether it emerged early or late. By contrast, decoupling of the second strongest fuel ($F_2$) frequently appeared transiently but disappeared before the final generations, whereas decoupling of the weakest fuel ($F_3$) was typically short-lived and indistinguishable from random topological fluctuations (Fig.~\ref{fig:fig3}c). These results indicate that evolutionary selection strongly discriminates among fuels according to their energetic contribution.

We further tested the causal role of fuel identity by perturbing evolved networks through targeted recoupling of specific fuels. Starting from evolved topologies obtained in the simulations, we selectively reintroduced individual fuel--substrate coupling reactions and quantified the resulting changes in dissipation and functional performance (Fig.~\ref{fig:fig3}d). Recoupling lower-energy fuels ($F_2$ or $F_3$) produced little effect on either dissipation or performance. In sharp contrast, recoupling the high-energy fuel ($F_1$) led to a dramatic reduction in dissipation and a pronounced loss of functional accuracy. These perturbation experiments highlight the dominant role of decoupling the highest-energy fuel in sustaining non-equilibrium dissipation and network performance.

\subsection{Fuel decoupling is a necessary condition for strong non-equilibrium driving}

\begin{figure}%[tbhp]
	\centering
	\includegraphics[width=.8\linewidth]{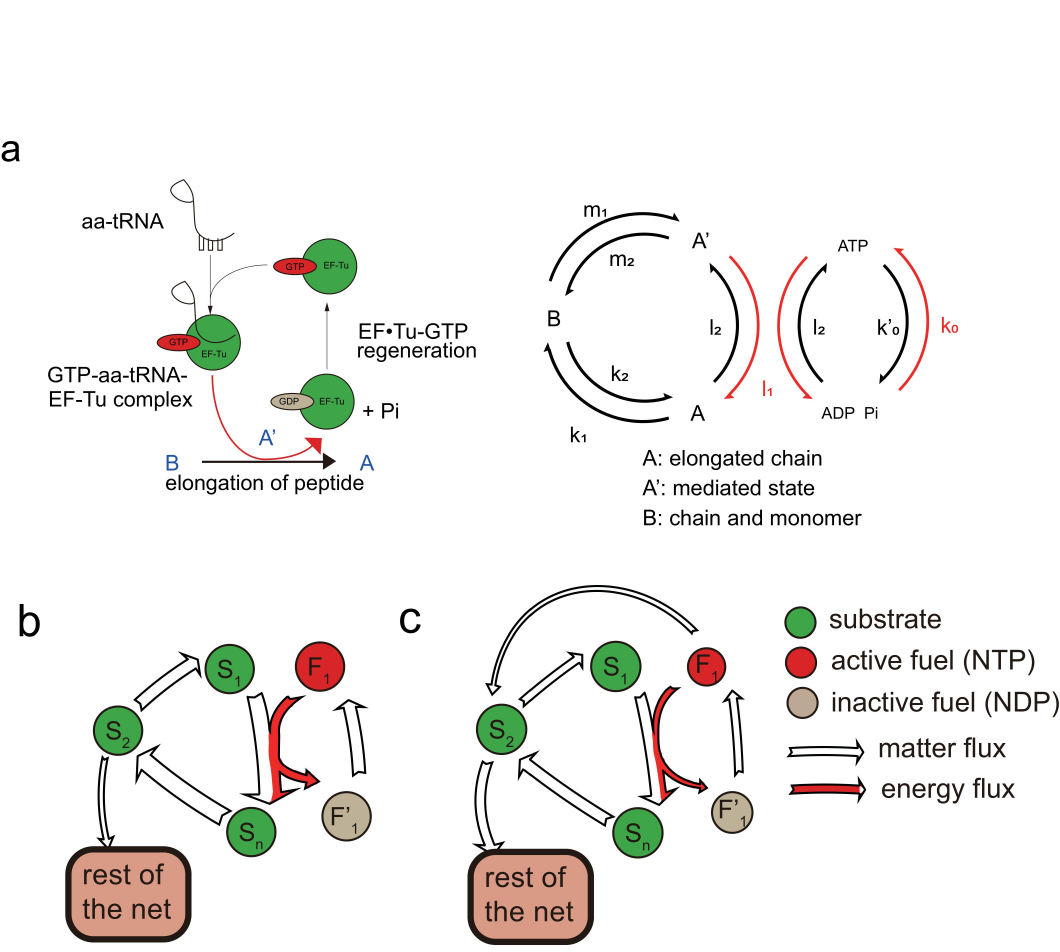}
	\caption{
	\textbf{Schematic illustration of theoretical analysis of fuel decoupling in the proofreading networks.}
	\textbf{(a)} Simplified reaction scheme based on Hopfield’s kinetic proofreading model~\cite{hopfield1974}. $B$ represents the chain and separated monomer, which can be viewed as a state at the beginning of elongation. The driving reaction occurs on the conversion $A' \rightarrow A$, analogous to energy-coupled elongation in biological systems. ATP hydrolysis (red arrows) provides the energy that maintains the system far from equilibrium. Model parameters $m_1$ and $k_2$ represent effective kinetic constants incorporating substrate concentration. 
	\textbf{(b,c)} Schematic illustrations of decoupled \textbf{(b)} and coupled \textbf{(c)} fuel--substrate configurations. Arrow width indicates the magnitude of matter flux in each direction, with red segments denoting flux driven by the high-energy fuel ($F_1$). In the decoupled case \textbf{(b)}, the driven flux is substantially larger than in the coupled case \textbf{(c)}, resulting in a higher non-equilibrium matter flow consistent with analytical predictions.
	}
	\label{fig:fig4}
\end{figure}

Having shown that decoupling the strongest fuel is selectively favored during evolutionary simulations, we asked whether this preference reflects a deeper physical constraint: is fuel decoupling merely advantageous, or is it required to sustain strong non-equilibrium driving in functional networks? We considered a simplified proofreading scheme based on Hopfield’s classical model, in which monomers bind to an elongating chain ($B$) to form an intermediate complex ($A'$), which is subsequently converted into the elongated chain ($A$), driven by an external fuel $F$ (Fig.~\ref{fig:fig4}a). At steady state, the concentration of state $A$, $[A]$, can be expressed as a function of kinetic parameters and $[B]$ (concentration of state $B$), as shown in Eq.~(1). Following the formulations of Hopfield~\cite{hopfield1974} and Qian~\cite{qian2006}, the difference between correct and incorrect pathways arises from the distinct dissociation rates $k_1$ and $m_2$. Therefore, we can rewrite these two parameters for the incorrect product as $m_2' = m_2 e^{dE}$ and $k_1 = k_1' e^{dE}$, where $dE$ is the energy difference between the correct and incorrect product. Assuming $dE$ is small, we rewrite the performance of the network as Eq.~(2).

\begin{equation}
[A]
=
\frac{
m_1 l_1 + k_2 l_1 + k_2 m_2 + (m_1+k_2)k_d[F]
}{
m_2 l_2 + k_1 l_1 + k_1 m_2 + m_2 k_d'[F'] + k_1 k_d[F]
}
[B]
\label{eq:1}
\end{equation}

\begin{equation}
s
=
\frac{1}{1+[A]_{\mathrm{incorrect}}/[A]_{\mathrm{correct}}}
=
\left(
2-dE-dE
\left(
\frac{k_1m_2}{
k_1m_2+m_2k_d'[F']+k_1k_d[F]
}
\right)
\right)^{-1}
\label{eq:2}
\end{equation}

We define the matter flux
$
M=l_2k_d[F]-l_1k_d'[F']
$
and the non-equilibrium level of the kinetic parameter
$
\gamma=[F']/[F]
$.
Substituting these definitions gives

\begin{equation}
s
=
\left(
2-dE-dE
\left(
\frac{k_1m_2}{
k_1m_2
+
m_2\left(\frac{k_d'M}{a\gamma-b}\right)
+
k_1\left(\frac{k_dM\gamma}{a\gamma-b}\right)
}
\right)
\right)^{-1}
\label{eq:3}
\end{equation}

where $a=l_2k_d$ and $b=l_1k_d'$.

Eq.~\ref{eq:3} highlights a fundamental trade-off: while non-equilibrium driving ($\gamma$) enhances discrimination, excessive matter flux through the functional motif ($M$) degrades performance (Supplementary Note~3).

We now examine how fuel decoupling affects these two quantities. When a fuel molecule directly participates in substrate interconversion reactions, the matter flux through the functional motif necessarily increases. Because fuels typically occupy high-energy states and are maintained at elevated concentrations by external input~\cite{traut1994}, direct coupling leads to an increase in total substrate concentration through Arrhenius scaling. As a result, matter flux grows, pushing the system toward a regime where functional precision is reduced (Fig.~\ref{fig:fig4}c).

More importantly, direct fuel coupling imposes a strict upper bound on the non-equilibrium parameter $\gamma$. In the fuel-coupled architecture, the maximum achievable non-equilibrium driving saturates at a value determined by the total amount of matter within the functional cycle, regardless of how strongly the fuel is driven externally (Eq.~(5), Fig.~\ref{fig:fig4}c and Supplementary Note~3). Conversely, when fuels are topologically decoupled from substrate exchange, the non-equilibrium parameter $\gamma$ monotonically approaches its theoretical maximum $k_1/k_0$ as the total concentration $[F]+[F']$ increases (Eq.~(4), Fig.~\ref{fig:fig4}b). Thus, fuel decoupling fundamentally removes the constraint on how far the system can be driven from equilibrium.

The following two equations describe the limiting behavior of $\gamma$ in the fuel-decoupled and fuel-coupled architectures, respectively. For the fuel-decoupled architecture, we have
\begin{equation}
\gamma
=
\frac{[F']}{[F]}
=
\frac{
k_1+d_+a\frac{N_t}{c+d[F']+e[F]}
}{
k_0+d_-a'\frac{N_t}{c+d[F']+e[F]}
}
\rightarrow
\frac{k_1}{k_0}.
\label{eq:4}
\end{equation}

\noindent
For the fuel-coupled architecture, we have
\begin{equation}
\gamma
=
\frac{[F']}{[F]}
=
\frac{
k_1+d_+a\frac{N_t+[F']+[F]}{c+d[F']+e[F]}
}{
k_0+d_-a'\frac{N_t+[F']+[F]}{c+d[F']+e[F]}
}
\rightarrow
\frac{
k_1+d_+a\frac{[F']+[F]}{d[F']+e[F]}
}{
k_0+d_-a'\frac{[F']+[F]}{d[F']+e[F]}
}
<
\frac{k_1}{k_0}.
\label{eq:5}
\end{equation}

The parameters satisfy
$\frac{d_-}{d_+} = \frac{e^{-E(S_d,F)}}{e^{-E(S_d',F')}}$,
$\frac{a}{a'} = \frac{e^{-E(S_d)}}{e^{-E(S_d')}}$,
and
$\frac{k_0}{k_1} = \frac{e^{-E(F)+E_{\mathrm{out}}}}{e^{-E(F)}}$,
where $S_d$ and $S_d'$ are substrates in the driving reaction and $E_{\mathrm{out}}$ is the external energy input converting $F'$ into $F$ (e.g., ADP to ATP). Constants $c$, $d$, and $e$ are network parameters, and $N_t$ is the total amount of matter in the system.

Because functional performance $s$ depends positively on the non-equilibrium parameter $\gamma$ (Eq.~\ref{eq:3}), this constraint implies that fuel-coupled networks are intrinsically limited in achievable accuracy. No adjustment of kinetic parameters can compensate for this limitation. Fuel decoupling is therefore a necessary topological condition for sustaining high non-equilibrium driving and, consequently, high functional precision in proofreading networks.

This theoretical conclusion extends beyond proofreading. In a broad class of non-equilibrium biochemical systems, including negative feedback loops, incoherent feed-forward loops and oscillatory circuits, functional performance similarly increases with non-equilibrium driving~\cite{cao2015,qian2006}. Because Eqs.~\ref{eq:4} and \ref{eq:5} do not depend on specific network topology, they are expected to apply generally to these systems. As a result, the parameter $\gamma$ is subject to the same constraints as in proofreading networks, thereby limiting achievable performance. Taken together, these results establish fuel decoupling as a general topological requirement for strong non-equilibrium function, providing a mechanistic explanation for its consistent emergence in evolutionary simulations.

\subsection{Fuel decoupling increase system robustness along evolution trajectories}

\begin{figure}%[tbhp]
	\centering
	\includegraphics[width=.8\linewidth]{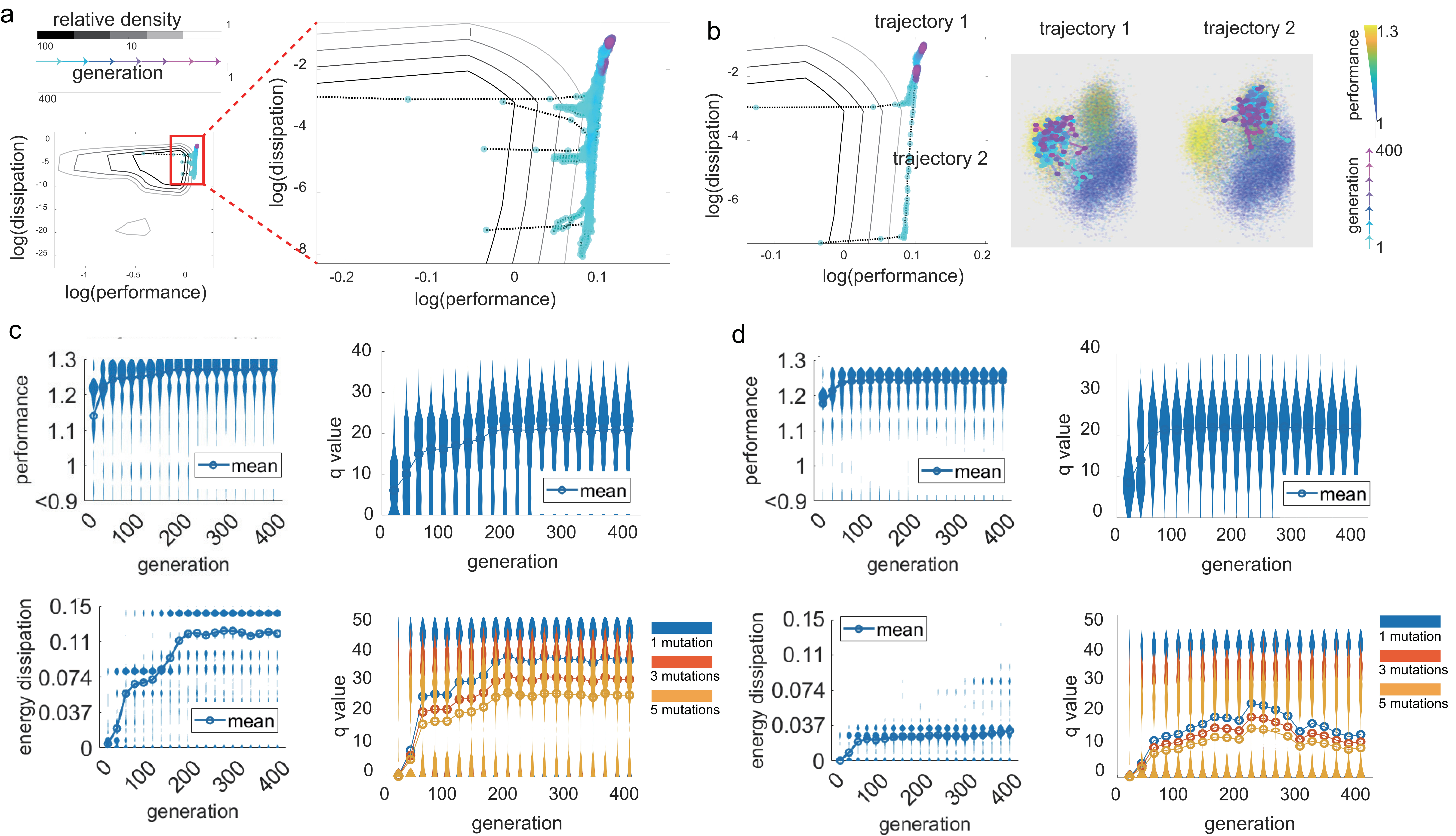}
	\caption{
	\textbf{Evolutionary trajectories and robustness analysis.}
	\textbf{(a)} Evolutionary trajectories of the proofreading model plotted in the performance--dissipation space. Contours indicate the density distribution of randomly generated networks. Each dot represents the mean position of 500 networks per generation, with the generation index indicated by the color bar. The left panel shows the global view, and the right panel shows a zoomed-in view of the high-performance and high-dissipation region enclosed in the red box. 
	\textbf{(b)} Evolutionary trajectories in topology space. Left panel shows the selected two trajectories from \textbf{(a)}. Right panel shows the corresponding paths in topology space. Each node represents a distinct network topology encountered during simulations, and edges correspond to single structural mutations. The distance between nodes reflects topological difference. Network performance and generation indices are indicated by the color bars. 
	\textbf{(c,d)} Quantitative analysis of Trajectories~1 \textbf{(c)} and 2 \textbf{(d)}. Each figure displays four performance metrics---functional performance (upper left), energy dissipation (lower left), parameter stability (upper right), and structural stability (lower right)---tracked across generations. Open circled lines represent the mean values for each case.
	}
	\label{fig:fig5}
\end{figure}

We next examined how fuel decoupling, as a general topological feature, influences evolutionary dynamics and network robustness. To this end, we tracked the evolutionary trajectories of proofreading networks in the joint space of functional performance and energy dissipation, where each position possesses a local density of topologies---defined as the topological entropy of that state---while selective pressure imposes an effective energy field along the performance axis (Fig.~\ref{fig:fig5}a and Methods). Initially, randomly generated networks (light blue dots) occupied regions of highest topological entropy, enclosed by black contours. As evolution proceeded under selection for proofreading accuracy, network populations consistently migrated rightward, following a directed trajectory toward regions characterized by both high performance and high dissipation (magenta dots in Fig.~\ref{fig:fig5}a). Notably, these regions correspond to low-density areas of the landscape that are rarely accessible through random sampling (white to light grey contours). Despite their entropic scarcity, evolved populations reproducibly converged there across independent simulations, indicating the presence of strong evolutionary attractors generated by functional selection.

To relate this behavior to changes in network topology, we projected two representative evolutionary trajectories from Fig.~\ref{fig:fig5}a into the underlying topology space, where nodes represent distinct network topologies and edges correspond to single topological mutations (Fig.~\ref{fig:fig5}b). Because networks occupying similar positions in the performance--dissipation landscape can nevertheless differ substantially in microscopic topology, highly functional and dissipative states form a densely connected cluster within the broader topology space. This clustering indicates that once fuel decoupling is established, numerous neighboring topologies preserve similarly high performance, suggesting enhanced structural resilience.

Comparison of individual trajectories revealed distinct evolutionary routes toward this clustered region (Fig.~\ref{fig:fig5}b). Trajectory~1 represents a rapid evolutionary process in which the population enters the high-performance cluster early and remains there throughout subsequent evolution. In contrast, Trajectory~2 represents a slower process in which the population traverses extended regions of intermediate performance before eventually converging onto the same cluster. Despite these differences in evolutionary timing, all successful trajectories shared a common structural feature: progressive decoupling of fuel molecules from the core functional motif. This convergence suggests that fuel decoupling defines the attractor itself, rather than the specific evolutionary path leading toward it.

To quantify robustness along evolutionary trajectories, we measured both parameter robustness and topological robustness using a $q$-value metric~\cite{ma2009}, defined as the number of perturbed topologies that retain functional performance above a fixed threshold following small perturbations to a focal topology. For each network encountered during evolution, we performed 50 replicate simulations with random perturbations to kinetic parameters or network topology (1, 3, or 5 mutations), followed by numerical evaluation of functional performance (Fig.~\ref{fig:fig5}c,d). Robustness against both classes of perturbations increased systematically over evolutionary time.

Importantly, differences in robustness between trajectories could be directly related to their position within the topological cluster. Trajectory~1, which entered the fuel-decoupled cluster earlier, exhibited substantially greater structural robustness, whereas Trajectory~2, which approached the cluster more gradually, displayed reduced tolerance to topological perturbations. These results indicate that fuel decoupling enhances not only functional performance but also resilience against random genetic perturbations.

Together, these analyses demonstrate that fuel decoupling fundamentally reshapes the evolutionary landscape. By enabling strong non-equilibrium driving, decoupling guides evolving populations toward rare yet densely connected regions of topology space that simultaneously combine high performance with enhanced robustness. In this sense, fuel decoupling transforms energetic specialization into an evolutionary advantage, generating stable attractors that promote both functional efficiency and long-term evolvability.

\subsection{Fuel decoupling emerges in coupled multifunctional systems}

\begin{figure}%[tbhp]
	\centering
	\includegraphics[width=.8\linewidth]{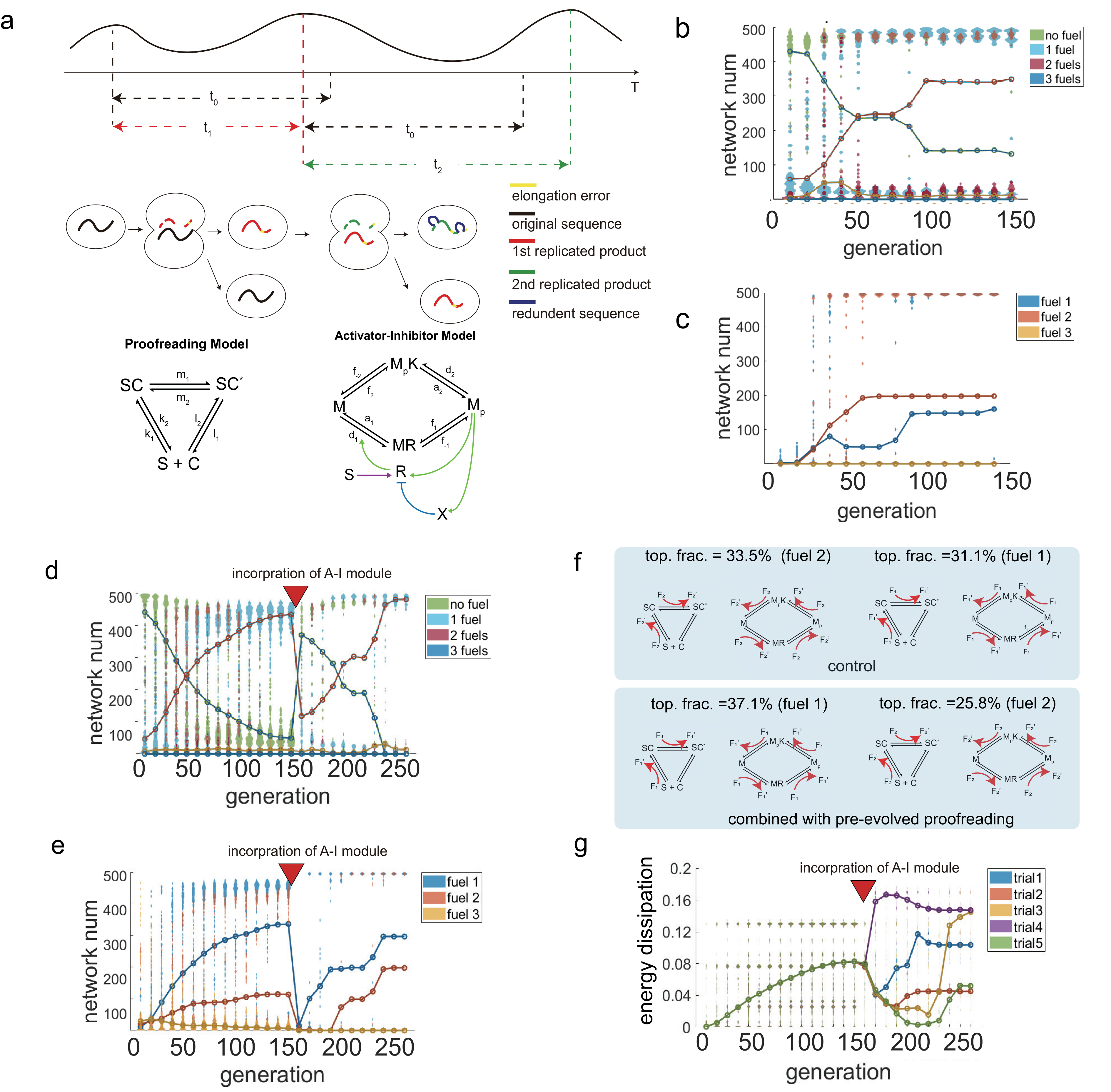}
	\caption{
	\textbf{Fuel decoupling and modular integration in the timed replicator model system.}
	\textbf{(a)} Schematic of the timed replicator model for elongation and replication. The duration of each replication cycle is controlled by the activator--inhibitor (A--I) oscillator, whereas replication accuracy is determined by the proofreading module embedded within the system. Representative core structures of the proofreading and oscillator modules are shown. 
	\textbf{(b--c)} Evolution of fuel decoupling states in 5 independent simulations of the timed replicator model. Open circled lines represent the mean number of networks in each group. 
	\textbf{(d--e)} Evolution of fuel decoupling states with integration of the randomly initialized A--I oscillator into the proofreading module (evolved to generation 150, indicated by the red arrow). Open circled lines represent the mean number of networks in each group. 
	\textbf{(f)} Evolved topologies under different integration strategies. Upper panel shows the proportion of fuel-decoupled topologies in the control case (with both modules evolved from scratch), while lower panel shows the case with pre-evolved proofreading module. 
	\textbf{(g)} Energy dissipation trajectories after incorporation of the A--I oscillator into the pre-evolved proofreading module. Open circled lines represent the mean dissipation of networks for each trial.
	}
	\label{fig:fig6}
\end{figure}

Biological networks rarely operate as isolated functional motifs; instead, multiple non-equilibrium processes are typically coordinated to jointly determine organismal fitness. To address whether fuel dedication and decoupling also emerge at the systems level, and how energetic resources are allocated during evolution to drive distinct subnetworks, we devised a toy model of a bi-functional self-replicating system---the timed replicator model (Fig.~\ref{fig:fig6}a and Supplementary Note~1). The model integrates two coupled motifs: (1) a kinetic proofreading network ensuring faithful replication of an information polymer, and (2) an activator--inhibitor oscillator functioning as a timing mechanism controlling replication duration.

In this system, replication accuracy depends on the proofreading module; insufficient proofreading introduces replication errors, modeled as additively deleterious mutations. Meanwhile, the oscillator determines when replication is terminated and information is segregated. Fluctuations in the timing module lead either to information loss through premature termination or to redundant information accumulation through delayed termination, assuming a constant replication speed.

This system represents an abstraction of a synthetic cell coupling DNA replication with a primitive cell cycle to achieve the core function of faithful information inheritance. Successful replication requires that a sufficient fraction of correct information be generated before the timing module triggers information segregation. The overall fitness of the system was therefore determined by two components. First, the proofreading accuracy,
$
s_{\mathrm{err}}
=
(L_{\mathrm{correct}}C^*)
/
(L_{\mathrm{correct}}C^*
+
L_{\mathrm{wrong}}C^*
+
\sum_i F_iC^*)
$.
Second, the ratio of faithfully inherited information,
$
P_{\mathrm{red}}/P_{\mathrm{green}}
$,
after a fixed number of replication cycles (Fig.~\ref{fig:fig6}a). Although dissipation has previously been shown to improve both oscillatory stability and replication fidelity in isolated systems, its influence on the coupled functional output of composite dissipative systems has remained unclear.

As in previous simulations, the system was initialized with the core oscillator and proofreading motifs together with three randomly connected fuel molecules differing in energetic strength. Network topology was allowed to evolve through stochastic mutations, while selection acted solely on the composite functional output without directly constraining energy dissipation or fuel usage (Supplementary Note~1 and 2). Despite the increased structural complexity and coupling between functions, evolutionary trajectories consistently exhibited the same qualitative pattern of fuel decoupling (Fig.~\ref{fig:fig6}b,c and Fig.~S13). These results suggest that the principle of fuel decoupling extends beyond simple isolated motifs to multifunctional integrated systems.

Closer examination of this composite model further revealed that fuel dedication becomes less dominant when multiple functional motifs coexist (Fig.~\ref{fig:fig6}b,c). The proportion of networks decoupling $F_1$ became comparable to, or even lower than, the proportion decoupling $F_2$. However, analysis of static performance landscapes relating fuel-decoupling state to dissipation and functional performance showed that $F_1$-decoupled networks consistently achieved higher dissipation and superior overall function. This apparent discrepancy can be understood from the structure of the evolutionary landscape itself. For example, local basins enriched in $F_2$-decoupled topologies can transiently trap evolving populations, and escaping such basins becomes increasingly difficult when multiple modules are coupled together (Fig.~\ref{fig:fig5}b).

To test this interpretation, we incorporated proofreading subnetworks evolved up to generation~150 into the composite system (Fig.~\ref{fig:fig6}d--g). Subsequent evolution then shifted the population toward states of higher performance and stronger dissipation dominated by strong-fuel decoupling (Fig.~\ref{fig:fig6}f and Fig.~S14--S15). These results suggest that modular evolution facilitates efficient integration of multifunctional systems and that previously evolved subnetworks can promote fuel dedication in newly introduced modules.

Together, these results extend the principle of fuel decoupling and dedication from isolated network motifs to coupled multifunctional systems. Even under increased structural and functional complexity, sequential selection acting on composite functions drives the topological separation of fuels from multiple subnetworks, giving rise to the bipartite organization that characterizes dissipative biological networks.

\section{Discussion}

In this study, we investigated how reaction network topology evolves under selection for non-equilibrium biological function. Across diverse dissipative motifs, we consistently observed the spontaneous emergence of fuel decoupling during evolution. Importantly, this outcome arose without any explicit selection on energy dissipation, indicating that fuel decoupling is not an auxiliary optimization but a structural consequence of functional selection in driven systems.

This behavior provides a natural explanation for the pervasive use of dedicated energy currencies in biology. Although cells can harness energy from diverse sources, this energy is largely funneled into a small set of high-energy molecules such as ATP, typically present at millimolar concentrations, higher than most metabolic substrates~\cite{beard2008,traut1994,teusink1998}. Such abundance makes ATP an effective global energy currency, but it also makes direct fuel--substrate coupling potentially problematic: if ATP is embedded in regulatory reaction motifs, fuel flux and material flux would become entangled, limiting the attainable non-equilibrium driving. Consistent with this intuition, our simulation results and theoretical analysis show that direct fuel--substrate coupling imposes an upper bound on achievable non-equilibrium driving ($\gamma$) that cannot be overcome by tuning kinetic parameters. Fuel decoupling removes this constraint by separating energetic input from material flux, thereby providing a network-level form of energy channeling~\cite{zala2017}: free energy can be delivered through dedicated carrier cycles, while module-specific material fluxes remain insulated. This architecture allows sustained, high-level dissipation and thus higher functional accuracy. Meanwhile, it is also compatible with ATP's broader physiological roles, including proposed function as a molecular hydrotrope~\cite{greiner2021,patel2017}, because ATP can remain abundant without being directly mixed into module-specific substrate pools. From this perspective, decoupling is not merely advantageous but a necessary topological condition for non-equilibrium networks.

Beyond performance, fuel decoupling shapes evolutionary dynamics. Decoupled architectures guide populations toward densely connected regions of topology space characterized by enhanced robustness. These regions function as evolutionary attractors: once reached, networks can tolerate substantial parameter fluctuations and structural mutations without loss of function (Fig.~\ref{fig:fig5}). In this sense, energetic specialization is directly linked to evolvability. Decoupling not only stabilizes network dynamics but also promotes the accumulation and retention of functional innovations. This landscape perspective also offers a possible explanation for the persistence of secondary fuels such as GTP. In our simulations, decoupling of lower-energy fuels was generally less globally optimal but could appear within local basins of the evolutionary landscape. Biologically, GTP-dependent processes may represent analogous solutions that became stabilized in particular functional contexts where the energetic disadvantage relative to ATP is modest, or the local network architecture is already compatible with GTP usage. This view is consistent with the specialized roles of GTP in translation, cytoskeletal regulation, small GTPase signaling, and other processes where GTP functions not as a replacement for ATP, but as a secondary fuel associated with specific regulatory logic~\cite{mrnjavac2024}. Therefore, ATP dominance and GTP persistence are not contradictory: ATP represents the globally favored high-energy currency, whereas GTP-dependent pathways may reflect historically and topologically constrained local optima within a broader evolutionary landscape.

The persistence of fuel decoupling in coupled multifunctional systems further places this principle in the context of biological modularity. In our bi-functional model (Fig.~\ref{fig:fig6}), proofreading and oscillatory control were required to cooperate through shared energetic resources, yet evolution still favored the decoupled architecture of energetic input and converged toward the use of a single dominant fuel. This observation suggests that fuel decoupling may have facilitated the integration of independently optimized non-equilibrium motifs into larger cellular systems. By decoupling common energetic flux from module-specific reaction pathways, networks can preserve global energy homeostasis while allowing regulatory, informational, and mechanical submodules to evolve semi-independently. Such modularized evolution is a central feature of modern biology: cell-cycle timers, replication machineries, signaling cascades, and ATP-driven molecular machines each implement distinct functional logics, yet draw from common energetic currencies without being directly entangled with one another's material fluxes. In this sense, the bipartite organization provides a topological basis for the scalability of biological organization, enabling the integration of new functions, rewiring existing ones, and adapting to fluctuating environments. It also offers a natural route toward more hierarchical forms of energetic architecture, including compartmentalized systems such as mitochondria and chloroplasts, where dedicated energy-converting modules generate fuel molecules that support diverse downstream processes. Similar strategies are evident in natural systems such as metabolic compartmentalization and ATP-driven molecular machines~\cite{guzun2014,leighton2025}.

In conclusion, our analyses demonstrate that fuel decoupling and dedication constitute a general organizing principle of dissipative biological networks. By channeling energy through dedicated fuels while preserving the integrity of functional reaction motifs, this architecture helps explain why ATP dominates as the primary energy currency, while secondary fuels such as GTP can persist in specialized local context. More broadly, fuel decoupling may represent a fundamental strategy by which living systems manage the universal requirement of dissipation: energy is centralized, stabilized, and distributed through dedicated carriers, while functional modules remain sufficiently insulated from direct coupling to fuel, allowing integration into higher-order biological organization. This perspective may also inform future efforts to understand prebiotic and synthetic biochemical systems~\cite{lorenz2011,cuevas2023,gawthrop2014,wilkes2021,szejka2010}, not as a prescriptive rule, but as a physical and evolutionary logic underlying the emergence of robust, scalable living systems.

\section{Materials and Methods}

\subsection{Network models}

We investigated the evolutionary origin of fuel--substrate decoupling in dissipative biochemical networks by systematically analyzing and evolving the topologies underlying several canonical cellular functions, including kinetic proofreading, biochemical oscillation, biochemical adaptation, and artificial replication. Each model was constructed from previously established biochemical network frameworks defining the minimal reaction topology necessary to realize the corresponding biological function~\cite{hopfield1974,cao2015,lan2012}.

To introduce non-equilibrium driving and evolutionary variability, canonical networks were expanded by incorporating fuel-coupled reactions together with fuel--substrate transformation processes. Specifically, three classes of reactions were considered: (i) driving reactions coupling substrate transformations to fuel consumption, (ii) fuel--substrate interconversion reactions, and (iii) model-specific extension reactions allowed only in certain systems (e.g., fuel binding in kinetic proofreading networks). These reaction classes collectively generated an expanded space of dissipative network topologies while preserving the core functional structure of each model. Detailed reaction schemes and topological constraints are provided in Supplementary Note~1.

\subsection{Performance metrics}

For each network class, functionality was quantified using a model-specific performance score. In kinetic proofreading networks, performance was defined as the fraction of correctly activated products while accounting for fuel-induced incorrect activation. In oscillatory systems, performance was quantified by the phase diffusion coefficient, which characterizes oscillatory stability. In adaptation networks, performance was defined using a combined metric incorporating both sensitivity and precision of the adaptive response. In artificial replicator systems, performance integrated replication fidelity together with effective replication length.

All performance measures were normalized to permit evolutionary selection toward improved function. Precise mathematical definitions of all performance metrics are provided in Supplementary Note~1.

\subsection{Evolutionary simulations}

Network evolution was simulated using a mutation--selection framework operating directly on network topology. Starting from canonical ancestral networks, reaction edges corresponding to allowed reaction classes were randomly added or removed subject to predefined structural constraints. At each generation, network performance was evaluated under stochastic chemical kinetics, and selection favored networks exhibiting higher functional scores.

Multiple independent evolutionary trajectories were simulated for each network class. Network topology, functional performance, energy dissipation, and fuel--substrate connectivity were recorded throughout evolution. The evolutionary protocol is illustrated schematically in Supplementary Note~2 and Fig.~S5, while representative outputs are summarized in Fig.~S6 and Fig.~S7.

\subsection{Energy dissipation and fuel decoupling analysis}

Energy dissipation was quantified by computing the steady-state entropy production rate associated with fuel-driven reactions within the principal functional loop of each network. Fuel--substrate decoupling was assessed by determining whether fuels directly participated in substrate-binding reactions or instead remained topologically separated from the core functional module.

The emergence and persistence of fuel decoupling were tracked across evolutionary trajectories, and statistical analyses were performed across independent simulations to quantify general trends. Detailed definitions of dissipation measures and decoupling metrics are provided in Supplementary Note~1 and 2.

\subsection{Stability analysis}

To evaluate robustness, evolved networks were subjected to both parameter perturbation and structural mutation tests. Parameter robustness was quantified by randomly sampling kinetic parameters and measuring the fraction of parameter sets preserving performance above a predefined threshold. Structural robustness was assessed by introducing random reaction-level mutations and measuring retention of function.

Together, these analyses quantify both parametric and topological robustness of evolved networks. Full procedures and theoretical interpretations are described in Supplementary Note~2.

\subsection{Numerical implementation}

All simulations and data analyses were implemented in MATLAB (MathWorks).

For the activator--inhibitor oscillator model, chemical kinetics were simulated using the Gillespie stochastic simulation algorithm, which is required to quantify intrinsic noise and compute the phase diffusion coefficient characterizing oscillatory stability. For the kinetic proofreading, negative feedback loop (NFBL), and incoherent feed-forward loop (IFFL) models, system dynamics were described by deterministic ordinary differential equations (ODEs), and network performance was evaluated using steady-state solutions obtained through numerical integration until convergence.

Data visualization and statistical analyses were performed using custom MATLAB scripts. All simulation codes used in this study are publicly available via GitHub:
\url{https://github.com/s616442197-dotcom/fuel_decoupling_code.git}

\section{Acknowledgments}

This work was funded by the National Key Research and Development Program of China (2023YFF1206100).
\bibliographystyle{unsrt}  

\bibliography{modified_ref}

\end{document}